\documentclass[preprint,proceedings]{rmaa}

\SetYear{2002}
\SetConfTitle{Galaxy Evolution: Theory and Observations}

\title{Interpreting Offsets from the Tully-Fisher Relation}

\author{S. J. Kannappan\altaffilmark{1}, E. Barton Gillespie\altaffilmark{2}, D. G. Fabricant\altaffilmark{3}, M. Franx\altaffilmark{4}, \& N. P. Vogt\altaffilmark{5}}

\altaffiltext{1}{The University of Texas at Austin, 2511
Speedway, RLM 15.308, Austin, TX 78712, USA; sheila@astro.as.utexas.edu.}
\altaffiltext{2}{University of Arizona.}
\altaffiltext{3}{Smithsonian Astrophical Observatory.}
\altaffiltext{4}{Leiden Observatory.}
\altaffiltext{5}{New Mexico State University.}

\suppressfulladdresses

\listofauthors{S. J. Kannappan, E. Barton Gillespie, D. G. Fabricant, M. Franx, \& N. P. Vogt}
\indexauthor{Kannappan, S. J., Barton Gillespie, E., Fabricant, D. G., Franx, M., \& Vogt, N. P.}

\begin{document}
\maketitle 

\boldabstract{We have previously demonstrated that Tully-Fisher (TF)
residuals correlate with tracers of star formation history (color and
emission line equivalent width, EW) for a broad sample of Sa--Sd spiral
galaxies (Kannappan, Fabricant, \& Franx 2002, KFF).  Here we use these
correlations to study two other classes of galaxies: (1) galaxies in close
pairs, and (2) galaxies at intermediate redshift.}

\begin{figure}[!h]
\includegraphics[width=\columnwidth]{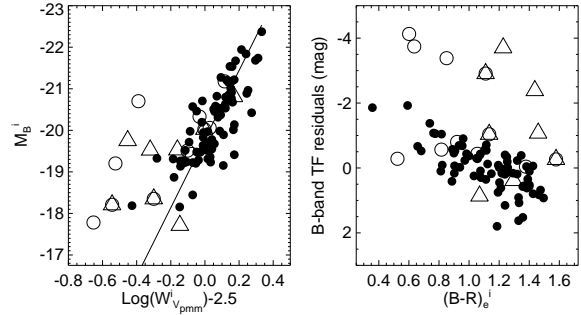}
\caption{TF relation and color--TF residual relation for galaxies in close
pairs.  Circles mark galaxies with RC asymmetries $>$10\% (see
KFF). Triangles mark galaxies with RCs extending to $<$0.9$r_e$.  The
remaining galaxies (solid points) define the inverse TF fit shown;
residuals are relative to this fit.}
\label{fig:pairs}
\end{figure}

In the close pairs sample of Barton et al.\ (2001), outliers
from the TF relation and the color--TF residual relation can be
independently identified based on their rotation curves (RCs): the
outliers' RCs nearly always show either strong asymmetry or severe radial
truncation (Figure~\ref{fig:pairs}).  Such RC abnormalities can affect
velocity widths, making it difficult to interpret TF residuals simply in
terms of luminosity offsets (star formation).  However, if we eliminate the
problem galaxies, we find a color--TF residual relation with a slope
similar to the slope defined by the Sa--Sd sample of KFF (drawn from the
Nearby Field Galaxy Survey, NFGS, Jansen et al.\ 2000).  The close pairs
relation extends further into the blue, reflecting recent interactions.

\begin{figure}[!t]
\includegraphics[width=\columnwidth]{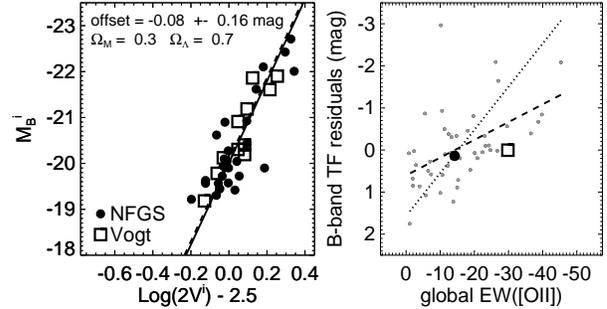}
\caption{(a) TF relations at low and intermediate $z$.  Squares show the
Vogt sample at $<$$z$$>$ $\sim$ 0.5, while large dots show a subsample of
the NFGS restricted in luminosity and morphology to match the Vogt
selection criteria.  (b) EW([OII])--TF residual correlation for bright
spirals in the NFGS (small dots).  Mean properties of the Vogt sample and
the Vogt-selected NFGS sample are overplotted (square and large dot).}
\label{fig:vogttfr}
\end{figure}

The intermediate-$z$ sample of Vogt et al.\ (1997) represents undisturbed
spiral galaxies brighter than M$_{\rm B}^i$ = $-$19.  Defining a similar
subsample within the NFGS and standardizing our analysis methods between
samples, we find no significant TF zero point evolution from $<$$z$$>$
$\sim$ 0.5 to $z$=0 (Figure~\ref{fig:vogttfr}a).  However, the
emission-line EWs in the Vogt sample are generally much higher than in the
NFGS, due to a combination of real evolution and detection bias in
emission-line fluxes.  Figure~\ref{fig:vogttfr}b shows that in the mean,
the Vogt galaxies deviate significantly from the $z$=0 EW--TF residual
relation, in the sense that for their EWs, they ought to have a brighter TF
zero point.  The same effect is seen in the color--TF residual relation.
Effectively, the Vogt galaxies are {\em underluminous} for their star
formation properties.  This result suggests that we may be seeing a
decrease in stellar mass fraction with lookback time, as less of the gas
has had time to form stars.



\begin{thebibliography}

%
%
%
%
\bibitem{b01} Barton, Geller, Bromley, 
van Zee, \& Kenyon. 2001, \aj, 121, 625

\bibitem{jffca} Jansen, Franx, Fabricant, \& Caldwell. 2000, 
	\apjs, 126, 271

\bibitem{kff} Kannappan, Fabricant, \& Franx. 2002, 
	\aj, 123, 2358

\bibitem{v97} {Vogt}, {Phillips}, {Faber},
	  {Gallego}, {Gronwall}, {Guzman},
	  {Illingworth}, {Koo}, \& {Lowenthal}.
	1997, \apjl, 479, L121

\end{thebibliography}
\end{document}